\documentclass{aa}  
\usepackage{graphicx}
\usepackage{txfonts}
\usepackage{amsmath}
\usepackage{xcolor}
\usepackage[colorlinks=true, linkcolor=blue, urlcolor=blue, citecolor=blue,]{hyperref}
\bibpunct{(}{)}{;}{a}{}{,} 
\begin{document}

   \title{On the origins of polarization holes in Bok globules}

   \author{R. Brauer\inst{1}
          ,
          S. Wolf\inst{1}
          \and
          S. Reissl\inst{2}
          }

   \institute{University of Kiel, Institute of Theoretical Physics and Astrophysics,
              Leibnizstrasse 15, 24118 Kiel, Germany\\
              \email{[rbrauer;wolf]@astrophysik.uni-kiel.de}
           \and
               University of Heidelberg, Institute of Theoretical Astrophysics,
               Albert-Ueberle-Str. 2 U04, 69120 Heidelberg, Germany\\
               \email{reissl@uni-heidelberg.de}
	      }


 
  \abstract
   {Polarimetric observations of Bok globules frequently show a decrease in the degree of polarization towards their central dense regions (polarization holes). This behaviour is usually explained with increased disalignment owing to high density and temperature, or insufficient angular resolution of a possibly complex magnetic field structure.}
   {We investigate whether a significant decrease in polarized emission of dense regions in Bok globules is possible under certain physical conditions. For instance, we evaluate the impact of optical depth effects and various properties of the dust phase.}
   {We use radiative transfer modelling to calculate the temperature structure of an analytical Bok globule model and simulate the polarized thermal emission of elongated dust grains. For the alignment of the dust grains, we consider a magnetic field and include radiative torque and internal alignment.}
   {Besides the usual explanations, selected conditions of the temperature and density distribution, the dust phase and the magnetic field are also able to significantly decrease the polarized emission of dense regions in Bok globules. Taking submm/mm grains and typical column densities of existing Bok globules into consideration, the optical depth is high enough to decrease the degree of polarization by up to $\Delta P {\sim}10\%$. If limited to the densest regions, dust grain growth to submm/mm size and accumulated graphite grains decrease the degree of polarization by up to $\Delta P {\sim}10\%$ and $\Delta P {\sim}5\%$, respectively. However, the effect of the graphite grains occurs only if they do not align with the magnetic field.}
   {}
 

  \maketitle

   \section{Introduction}

   The evolution of a collapsing molecular cloud to an evolved star is not yet fully understood. In particular, the impact of magnetic fields on star formation is an important part of the ongoing discussion \citep{matthews_magnetic_2002, pudritz_role_2014, seifried_impact_2015}. For instance, magnetic fields are considered as stabilising star-forming regions and filaments against contraction which is a possible explanation of observed low star formation rates, compared to the amount of gas in star-forming regions \citep{van_loo_magnetic_2015, federrath_inefficient_2015}. Magnetic fields are also expected to influence the shape of cloud fragments and the coupling between the gas and dust phase in molecular clouds \citep{henning_measurements_2001}. 

   In general, magnetic fields are indirectly measured by polarimetric observations in the near-IR or submm/mm regime \citep[see e.g.][]{bertrang_large-scale_2014}. Elongated dust grains are thought to align with their longer axis perpendicular to the magnetic field lines, which causes polarized thermal emission and dichroic extinction of the dust phase. Various mechanisms have been proposed to explain the alignment of dust grains (paramagnetic relaxation, \citealt{davis_polarization_1951}; mechanical alignment, \citealt{gold_alignment_1952}; radiative torque alignment, \citealt{lazarian_tracing_2007}). Nevertheless, the impact of each alignment mechanism strongly depends on the environment that is been considered. For example, mechanical alignment only appears in regions with supersonic flows \citep{lazarian_mechanical_1995} whereby radiative torque alignment is very efficient in a strong anisotropic radiation field \citep{matsumura_correlation_2011}.

   Polarimetric observations of molecular clouds sometimes show a decrease in the degree of polarization towards dense regions (polarization holes; \citealt{rao_high-resolution_1998}; \citealt{matthews_polarimetry_2005}). The origin of this effect is not yet understood, but is usually explained with an assumed increased disalignment of the dust grains, owing to higher density and temperature \citep{goodman_structure_1992, hildebrand_far-infrared_1999, henning_measurements_2001, wolf_magnetic_2003} or insufficient angular resolution of a possibly complex magnetic field structure \citep{glenn_magnetic_1999, wolf_evolution_2004, matthews_polarimetry_2005}. However, the relevance of these effects has not been confirmed quantitatively.

   In the optically thin case, elongated dust grains cause polarized emission that is due to the different absorption cross-sections of their axes. In the optically thick case, this difference of the cross-sections causes dichroic extinction instead. The transition between both cases is expected to show a decreased degree of polarization. For this reason, we investigate the influence of the optical depth and related physical properties of the dust on the polarized emission of Bok globules and their potential to significantly decrease the degree of polarization of dense regions.
   
   Given a sufficiently high angular resolution, all Bok globules have so far shown this quantitative behaviour of their radial submm polarization profile (\citealt{henning_measurements_2001}; \citealt{wolf_magnetic_2003}; \citealt{wolf_evolution_2004}). Bok globules are dense and cold molecular clouds where low-mass star formation typically takes place \citep{bok_small_1947, clemens_bok_1991}. They often have a simple spherical shape which supports modelling approaches. To achieve our aims, we create a generic analytical model of a Bok globule. We use observational constrained parameters of the Bok globule B335 to ensure that the model is a good approximation of a real Bok globule \citep{wolf_magnetic_2003, evans_b335:_2005, olofsson_new_2009, bertrang_large-scale_2014, launhardt_looking_2010}. The globule B335 is suitable because observations at $\lambda=850~\mathrm{\mu m}$ show that the degree of polarization decreases from $\sim\mathrm{15}\%$ at the envelope to $\sim\mathrm{2}\%$ at the centre. Furthermore, we vary selected parameters, such as dust grain size and density distribution, to cover different kinds of Bok globules and investigate the impact of such parameters on the answer of our key question.

   Based on our model of a Bok globule, we perform continuum radiative transfer simulations in the submm/mm regime with the radiative transfer code POLARIS (POLArized RadIation Simulator, {\color{blue}Reissl et al. subm.}) which is able to consider magnetic fields and the dust grain alignment mechanisms mentioned above. From POLARIS, we obtain intensity and polarization maps to investigate a potential decrease in the degree of polarization.

   We begin this study with an introduction of our reference model that is based on the Bok globule B335, which includes the chosen magnetic field and the alignment mechanisms (see Sect. \ref{model_description}). Subsequently, we present the results and start with an analysis of the influence of the optical depth on the degree of polarization in our reference model (see Sect. \ref{results}). This analysis is followed by a general discussion of the optical depth effect. After that, we investigate under which conditions the optical depth and other physical quantities are able to cause a significant decrease in the degree of polarization. Finally, we summarize the results and give a conclusion in Sect. \ref{conclusions}.

  \section{Model description}
  \label{model_description}

  \subsection*{Density distribution:}
  \label{density_distribution}

  The density distribution of the gas and dust phase is based on a Bonnor-Ebert sphere with a constant central density which can be written as follows \citep{kaminski_role_2014}:
  \begin{equation}
   \rho(r)=
    \begin{cases}
     \hspace{0.1cm} \rho_0 \cdot R_0^{-2}, \quad \text{if } r \leq R_0\\
     \hspace{0.1cm} \hspace{0.07cm}\rho_0 \cdot r^{-2}, \quad \text{if } R_0 < r \leq R_\mathrm{out}\\
     \hspace{0.1cm} \hspace{0.9cm}0, \quad \text{if } r > R_\mathrm{out}\\
    \end{cases}
    \label{eqn:density}
  \end{equation}
  Here, $\rho_0$ is a reference density used to achieve the preset Bok globule mass, $r$ is the radial distance from the central stellar object, and $R_\mathrm{0}$ is a truncation radius that defines the extent of the central region with constant density. 

  For our reference model, we consider a truncation radius of $R_\mathrm{0}=10^3~\mathrm{AU}$ and an outer radius of $R_\mathrm{out}=1.5\cdot10^4~\mathrm{AU}$ to fit with the radial brightness profile of the Bok globule B335 (see Fig. \ref{fig:wolf_B335}, left; \citealt{wolf_magnetic_2003}). Investigations of the line-of-sight optical depth need a model that is in agreement with column densities that are derived from existing Bok globules. Typical values of the beam-averaged hydrogen column density are between $N_\mathrm{H,beam}=10^{27}~\mathrm{m^{-2}}$ and $N_\mathrm{H,beam}=10^{28}~\mathrm{m^{-2}}$ \citep{launhardt_looking_2010}. This corresponds to total masses between $M_\mathrm{gas}\sim4~\mathrm{M_\odot}$ and $M_\mathrm{gas}\sim40~\mathrm{M_\odot}$ for our reference model which is in agreement with other mass estimations of Bok globules ($M_\mathrm{gas}=\{2,100\}~\mathrm{M_\odot}$; \citealt{bok_dark_1977}; \citealt{leung_physical_1985}; \citealt{clemens_bok_1991}). The default parameters of our reference model are summarised in Table \ref{tab:parameter}, but within the scope of our study, we vary selected parameters with regard to different kinds of Bok globules.

  \subsection*{Dust:}
  \label{dust}

  We assume compact, homogeneous and oblate dust grains, consisting of 62.5\% silicate and 37.5\% graphite (MRN-dust, \citealt{mathis_size_1977}; optical properties from \citealt{weingartner_dust_2001}). The ratio between the longer and shorter axis is $1:0.5$. For the grain size distribution we assume
   \begin{equation}
    \mathrm{d}n(a)\propto a^{-3.5} \text{d}a, \quad a_\mathrm{min} < a < a_\mathrm{max}, \label{eqn:dust}
   \end{equation}
  where $n(a)$ is the number of dust particles with a specific dust grain radius $a$. 

  Table \ref{tab:grain_growth} illustrates the spectral index of selected Bok globules that show one or more polarization holes. These as well as other Bok globules feature a low spectral index in the submm/mm regime of their total flux. A spectral index between two and three is usually a sign of advanced dust grain growth \citep{natta_dust_2007}. Considering dust grain growth up to submm/mm size in our reference model, the spectral index derived from the simulated flux at $850~\mathrm{\mu m}$ and $450~\mathrm{\mu m}$ amounts to $\sim2.8$, which is in agreement with observations (for minimum and maximum grain size see Table \ref{tab:parameter}). This spectral index corresponds to an emissivity index $\beta$ of $\sim0.8$.
  \vspace{0.05cm}
 
  \renewcommand{\arraystretch}{1.2}
  \begin{table}
  \caption{Overview of dust grain growth in selected Bok globules with one or more polarization holes. The spectral index of B335 is derived from the flux at $850~\mathrm{\mu m}$ and $450~\mathrm{\mu m}$, because flux measurements at $1.3~\mathrm{mm}$ are missing.}
  \label{tab:grain_growth}
  \centering
   \begin{tabular}{lll}
    \hline
    \hline
    Bok globule & Spectral index & Derived from $F_\lambda$ at\\
    \hline
    CB 54$^\mathrm{1,3}$ & 2.99 & $1.3~\mathrm{mm}$, $850~\mathrm{\mu m}$\\
    CB 199$^\mathrm{2,3}$ (B335) & 2.09 & $850~\mathrm{\mu m}$, $450~\mathrm{\mu m}$\\
    CB 224$^\mathrm{2,3}$ & 2.31 & $1.3~\mathrm{mm}$, $850~\mathrm{\mu m}$\\
    CB 230$^\mathrm{2,3}$ & 2.81 & $1.3~\mathrm{mm}$, $850~\mathrm{\mu m}$\\
    CB 244$^\mathrm{2,3}$ & 2.20 & $1.3~\mathrm{mm}$, $850~\mathrm{\mu m}$\\
    \hline
    \hline
   \end{tabular}\par\medskip
   \tablefoot{\footnotesize ($^1$\citealt{henning_measurements_2001}, $^2$\citealt{wolf_magnetic_2003}, $^{3}$\citealt{launhardt_looking_2010})}
  \end{table}
  \renewcommand{\arraystretch}{1}

  \subsection*{Magnetic field:}

  We assume a constant magnetic field strength of $B\sim140~\mathrm{\mu G}$ (B335, \citealt{wolf_evolution_2004}) in the whole Bok globule model. The magnetic field lines are oriented perpendicular with respect to the observer ($y$-direction in Figs. \ref{fig:mass_40} and \ref{fig:observation_test}). Other more complex magnetic field structures are also possible, but they would increase exceedingly the number of free parameters.

  \subsection*{Alignment mechanisms:}\label{align}

  We consider the following mechanisms for aligning oblate dust grains perpendicular to the magnetic field:
  \begin{itemize}
   \item \textbf{Radiative torque alignment (RAT)\\} For this mechanism, dust grains are considered as having a helicity that causes them to spin up by being radiated in an anisotropic radiation field with wavelengths less than their diameter \citep{lazarian_tracing_2007, lazarian_radiative_2007, hoang_radiative_2009, andersson_interstellar_2015}. The spinning dust grains align through the Barnett effect \citep{barnett_magnetization_1915}, which is opposed by gas bombardment, which reduces the alignment. We emphasise that our model uses perfect oblate spheroids to approximate oblate spheroids with a helicity. Usually, perfect oblate spheroids have no helicity.
   \item \textbf{Imperfect internal alignment\\} In general, the largest moment of inertia is not perfectly aligned with the angular momentum \citep{purcell_suprathermal_1979}. Fluctuations in the thermal energy will randomly excite rotation around all three body axes with energies in the order of $kT_\mathrm{d}$ ($T_\mathrm{d}:$ dust temperature). If the rotational energy of the grain is $\gg kT_\mathrm{d}$, then these thermal fluctuations are negligible and Barnett dissipation aligns the dust grains with their largest moment of inertia to the angular momentum \citep{jones_magnetic_1967, lazarian_barnett_1997}. However, the grains in molecular clouds are likely to have rotational energies $\sim kT_\mathrm{d}$ that results in imperfect internal alignment.
  \end{itemize}
  We do not consider paramagnetic relaxation \citep{davis_polarization_1951}, because simulations of our reference model show that this alignment mechanism is negligible compared to the RAT theory. We do also not assume supersonic flows and, therefore, we neglect mechanical alignment \citep{gold_alignment_1952}.

  \subsection*{Radiative transfer:}
  \label{radiative_transfer}
  We apply the three-dimensional continuum RT code POLARIS (POLArized RadIation Simulator, {\color{blue}Reissl et al. subm.}). This solves the radiative transfer problem self-consistently with the Monte Carlo method, and it considers magnetic fields as well as various dust grain alignment mechanisms. We use it to calculate the spatial temperature distribution of our Bok globule model and derive thermal emission maps using the full Stokes vector. The required optical properties of the dust grains are derived from the complex refractive index of the given material by using the Discrete Dipole SCATtering code (DDSCAT, see Sect. \ref{dust}; \citealt{draine_user_2013}). However, DDSCAT is not able to calculate the optical properties of dust grains with $a\gg\lambda$. We compensate for this by using the MIEX scattering routine for these dust grains \citep{wolf_mie_2004}. This approximation is justified, because their optical properties are almost isotropic at this size. We simulate the thermal emission of our reference model at $\lambda=850~\mathrm{\mu m}$, because many observations of polarization holes have been performed at this wavelength (e.g. with SCUBA/JCMT).

  The implementation of the radiative torque alignment requires us to specify the free parameter $f_\mathrm{high,J}$, which describes the ratio between dust grains rotating with high- and low-angular momentum. In the RAT theory, only these two rotational states are stable \citep{hoang_radiative_2008}. A grain with another angular momentum converges towards one of these stable points. The point of low angular momentum is prone to thermal bombardment of gas particles, which reduces the degree of alignment. Dust grains at the point of high angular momentum are not expected to be influenced by thermal disalignment and, therefore, perfectly aligned. In reality, $f_\mathrm{high,J}$ depends on physical quantities such as temperature and density but cannot be calculated analytically \citep{hoang_grain_2014}. In POLARIS, $f_\mathrm{high,J}$ is implemented by using one value for the whole model. For our reference model, we consider $f_\mathrm{high,J}=0$, but we also investigate the influence of other values in this parameter (see Sect. \ref{results}).

  \subsection*{Heating source:}\label{heating}
   
  The primary heating source is a central young stellar object (YSO). The YSO is characterised by an effective temperature and radius with values which are summarised in Table \ref{tab:parameter}.

   \renewcommand{\arraystretch}{1.2}
   \begin{table}
   \caption{Overview of the default parameters of our reference model. Values given for parameters shown with an apostrophe ($'$) are varied in the context of the parameter study.}
   \label{tab:parameter}
   \begin{tabular}{lll}
     \hline
     \hline
     \multicolumn{3}{c}{\textit{General}} \\
     \hline
   Distance to Bok globule &  $d$ & $100~\mathrm{pc}$ \\
   Observing wavelength & $\lambda$ & $850~\mathrm{\mu m}$ \\
     \hline
   \multicolumn{3}{c}{\textit{Central YSO}} \\
     \hline
   Effective temperature  &  $T$ & $6000~\mathrm{K}$ \\
   Radius  &  $R$ & $2~\mathrm{R_\odot}$ \\
     \hline
     \multicolumn{3}{c}{\textit{Density distribution}}\\
     \hline
   Truncation radius & $R_\mathrm{0}$ & $10^3~\mathrm{AU}$ \\
   Outer radius  &  $R_\mathrm{out}$ & $1.5\cdot10^4~\mathrm{AU}$ \\
     \hline
     \multicolumn{3}{c}{\textit{Gas phase}} \\
     \hline
   Gas mass  &  $M'_\mathrm{gas}$ & $8~\mathrm{M_\odot}$ \\
   Gas-to-dust ratio & $M_\mathrm{gas}:M_\mathrm{dust}$ & $100:1$ \\
     \hline
     \multicolumn{3}{c}{\textit{Dust phase}} \\
     \hline
   Fraction of silicate grains & $N'_\mathrm{silicate}$ & $62.5\%$\\
   Fraction of graphite grains & $N'_\mathrm{graphite}$ & $37.5\%$\\
   Silicate grain density  &  $\rho_\mathrm{silicate}$ & $3.8~\mathrm{g\ cm^{-3}}$\\
   Graphite grain density  &  $\rho_\mathrm{graphite}$ & $2.25~\mathrm{g\ cm^{-3}}$\\
   Minimum dust grain size  &  $a_\mathrm{min}$ & $5~\mathrm{nm}$ \\
   Maximum dust grain size  &  $a'_\mathrm{max}$ & $200~\mathrm{\mu m}$ \\
     \hline
     \multicolumn{3}{c}{\textit{Magnetic field}} \\
     \hline
   Magnetic field strength & $B$ & $134~\mathrm{\mu G}$\\
     \hline
     \multicolumn{3}{c}{\textit{Alignment parameter}} \\
     \hline
   Ratio of grains rotating at high $J$ & $f'_\mathrm{high,J}$ & $0$ \\
     \hline
   \end{tabular}
   \end{table}
   \renewcommand{\arraystretch}{1}

   \section{Results}
   \label{results}
   
  \subsection*{Reference model}
  \label{reference_model}
  
  Our reference model consists of a spherical distribution of perfectly mixed gas and dust around a stellar heating source in its centre (see Sect. \ref{heating}). The density distribution is given in Eq. \ref{eqn:density}; The size distribution of the dust grains is given in Eq. \ref{eqn:dust}. An overview of the parameters of the reference model is shown in Table \ref{tab:parameter}.
  
  In the following sections, we perform simulations using the reference model and vary only one particular parameter at a time to consider the behaviour of the degree of polarization in various Bok globules.
  
  We calculate the degree of polarization from the simulated thermal emission of the dust phase as follows:
  \begin{equation}
   P_\mathrm{l}=\frac{\sqrt{F_\mathrm{Q}^2+F_\mathrm{U}^2}}{F_\mathrm{I}}.
  \end{equation}
  Here, $F_\mathrm{Q}$ and $F_\mathrm{U}$ are the linear polarization components of the Stokes vector where $F_\mathrm{I}$ is the intensity component. The circular polarization component $F_\mathrm{V}$ is negligible in our scenarios, compared to the linear components and, therefore, not considered in subsequent discussions. Owing to the characteristics of the Stokes vector, the degree of polarization only provides information about the magnetic field components in the plane of sky (POS).

  \subsection*{Bok globule mass}
  \label{bok_mass}
\begin{figure*}[ht]
   \centering
   \includegraphics[width=0.9\hsize]{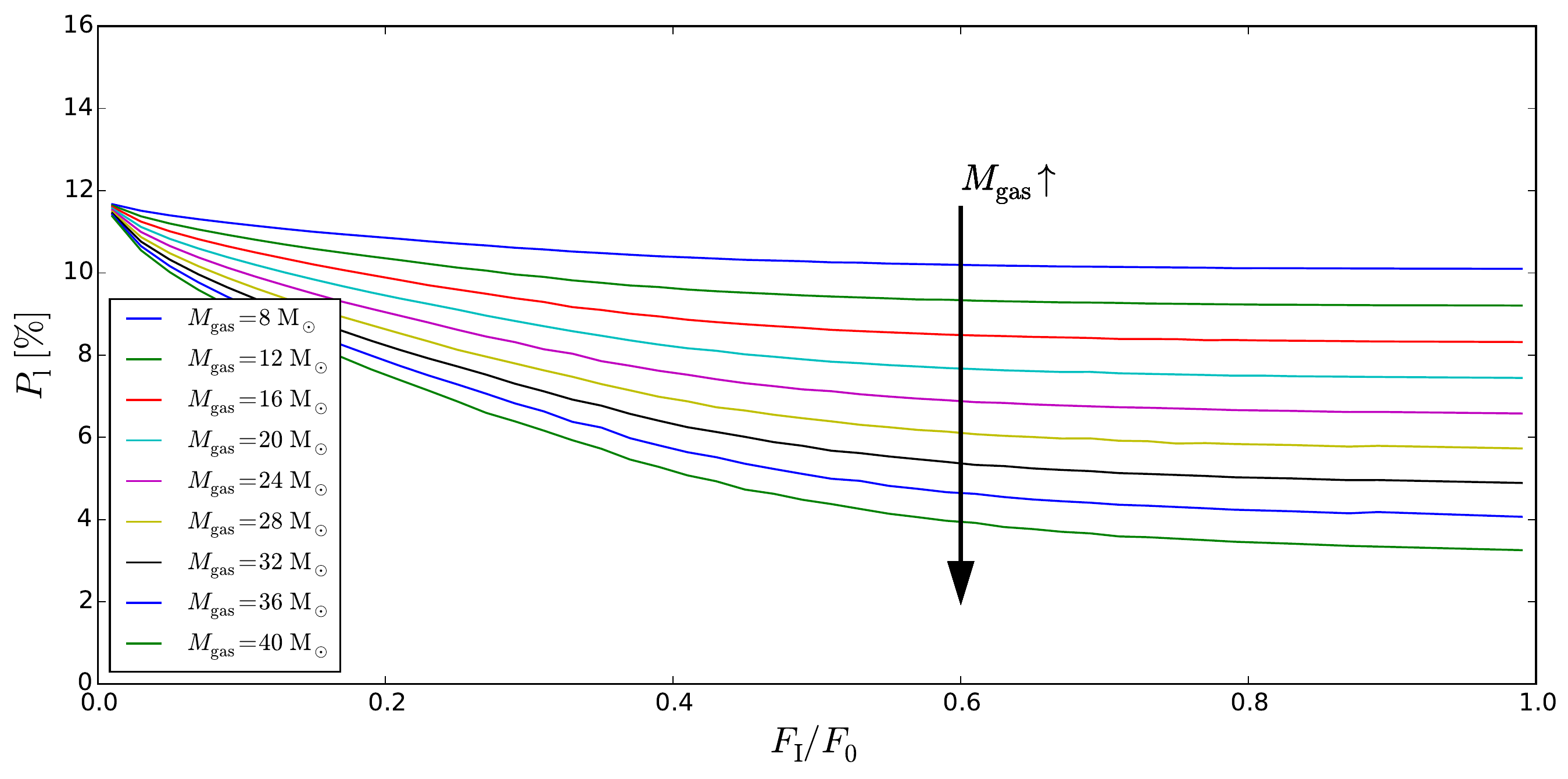}
   \caption{Distribution of the mean degree of polarization $P_\mathrm{l}$ vs. the intensity $F_\mathrm{I}$ across the Bok globule models. The intensity $F_\mathrm{I}$ is normalized to the maximum intensity $F_0$ of the respective Bok globule. The Bok globule models have total masses of $M_\mathrm{gas}=\{8,12,16,20,24,28,32,36,40\}~\mathrm{M_\odot}$ inside of $R=1.5\cdot10^4~\mathrm{AU}$ ($\lambda=850~\mathrm{\mu m}$).}
   \label{fig:mass_study}
  \end{figure*}

\begin{figure*}
   \centering
   \resizebox{0.9\hsize}{!}{\includegraphics[width=0.69\hsize]{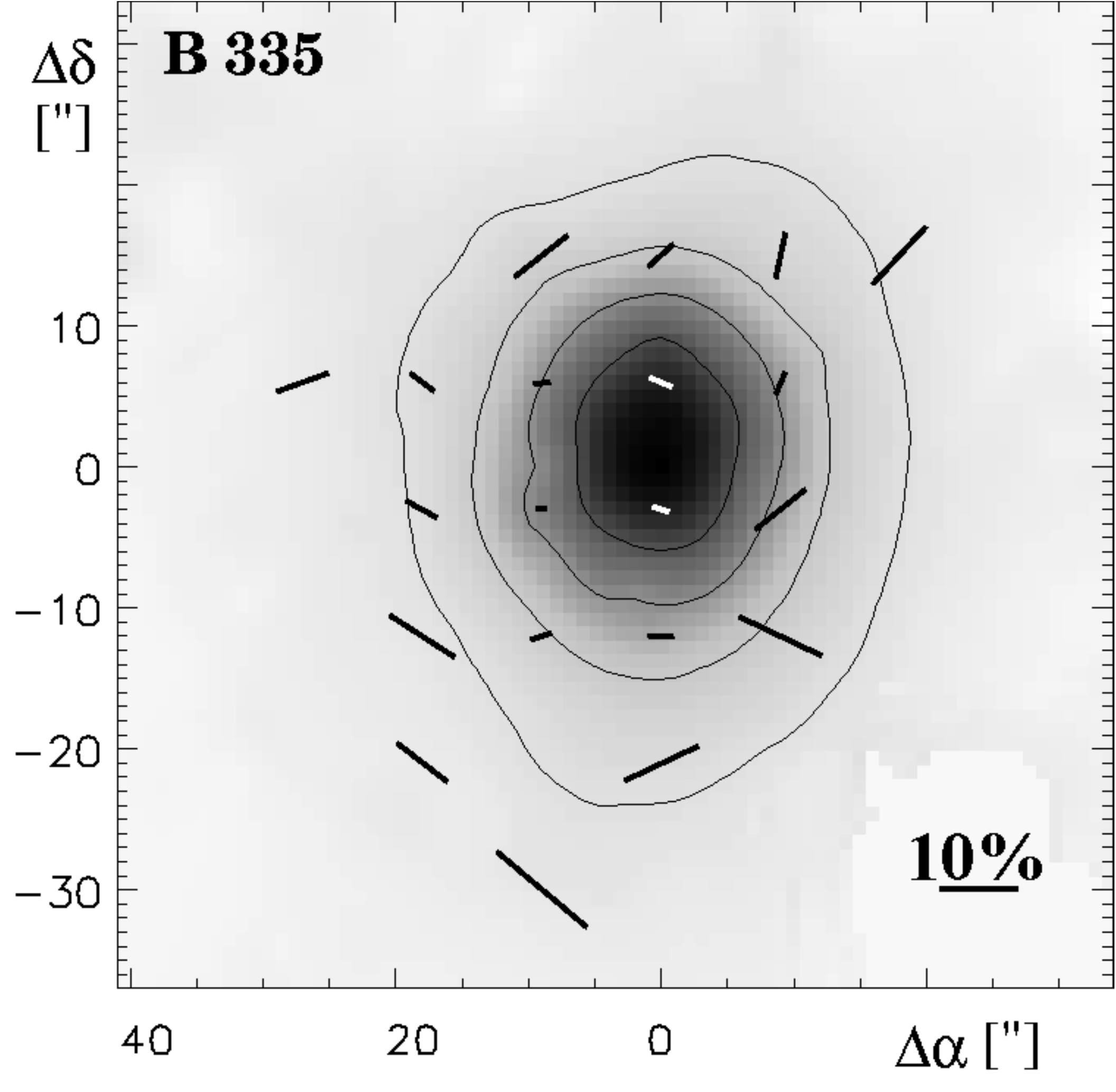} \quad \includegraphics[width=\hsize]{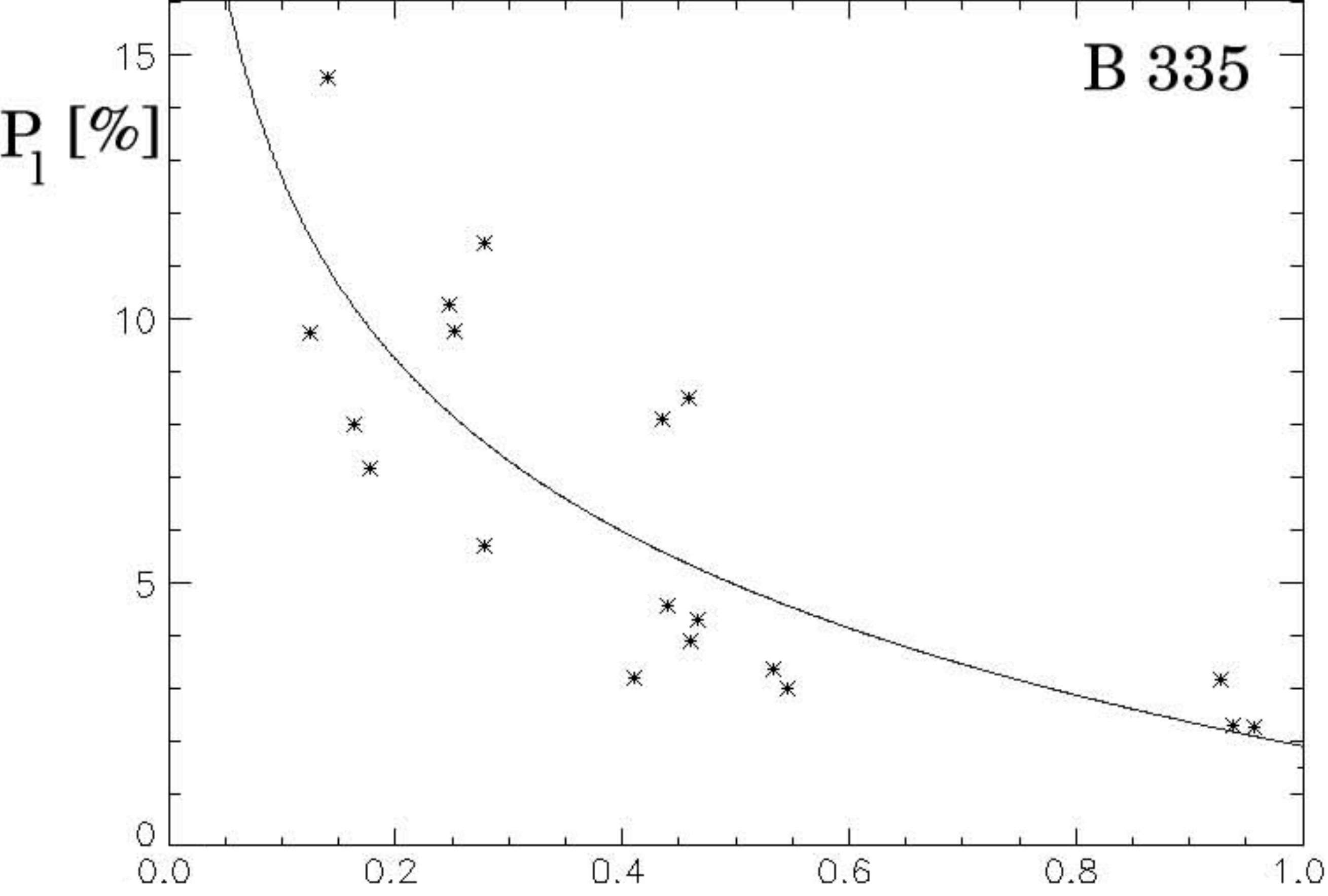}}
   \caption{\textit{Left:} SCUBA $850~\mathrm{\mu m}$ maps of B335 with polarization vectors superimposed. The length of the vectors is proportional to the degree of polarization and the direction gives the position angle. The data are binned over $9^{\prime\prime}$. Only vectors for which the $850~\mathrm{\mu m}$ flux exceeds 5 times the standard deviation and $P_\mathrm{l}/\sigma(P_\mathrm{l})>3$ are plotted. The contour lines mark $20\%$, $40\%$, $60\%$, and $80\%$ of the maximum intensity. \textit{Right:} Scatter diagrams showing the distribution of $P_l$ vs. intensity $I$ across the Bok globule B335. Only data points in which the $850~\mathrm{\mu m}$ flux exceeds 5 times the standard deviation and $P_l/\sigma(P_l) > 3$ have been considered. (\citealt{wolf_magnetic_2003}; \href{http://dx.doi.org/10.1086/375622}{DOI:10.1086/375622}; \textcopyright AAS. Reproduced with permission;)}
   \label{fig:wolf_B335}
  \end{figure*}

  We begin our investigations with the influence of the Bok globule mass and, therefore, the optical depth on the polarized emission of Bok globules. For this purpose, we simulate the thermal emission of our reference model with total masses of $M_\mathrm{gas}=\{8,12,16,20,24,28,32,36,40\}~\mathrm{M_\odot}$. These total masses correspond to the following optical depths:
  \begin{align*}
   M_\mathrm{gas}&=8~\mathrm{M_{\odot}}: \quad \tau_{850~\mathrm{\mu m}}=0.12\quad (A_V=10)\\
   M_\mathrm{gas}&=12~\mathrm{M_{\odot}}: \quad \tau_{850~\mathrm{\mu m}}=0.18\quad (A_V=15)\\
   M_\mathrm{gas}&=16~\mathrm{M_{\odot}}: \quad \tau_{850~\mathrm{\mu m}}=0.24\quad (A_V=20)\\
   M_\mathrm{gas}&=20~\mathrm{M_{\odot}}: \quad \tau_{850~\mathrm{\mu m}}=0.3\quad (A_V=25)\\
   M_\mathrm{gas}&=24~\mathrm{M_{\odot}}: \quad \tau_{850~\mathrm{\mu m}}=0.36\quad (A_V=30)\\
   M_\mathrm{gas}&=28~\mathrm{M_{\odot}}: \quad \tau_{850~\mathrm{\mu m}}=0.42\quad (A_V=35)\\
   M_\mathrm{gas}&=32~\mathrm{M_{\odot}}: \quad \tau_{850~\mathrm{\mu m}}=0.48\quad (A_V=40)\\
   M_\mathrm{gas}&=36~\mathrm{M_{\odot}}: \quad \tau_{850~\mathrm{\mu m}}=0.54\quad (A_V=45)\\
   M_\mathrm{gas}&=40~\mathrm{M_{\odot}}: \quad \tau_{850~\mathrm{\mu m}}=0.6\quad (A_V=50).
  \end{align*}
  Here, the optical depth is derived for the distance between the centre and the boundary of our Bok globule model ($\lambda=850~\mathrm{\mu m}$). Figure \ref{fig:mass_study} illustrates the degree of polarization vs. the normalized intensity. Within the chosen mass range, the optical depth effect decreases the degree of polarization by about $2\%$ to $10\%$. Compared with observations, this decrease amounts to about $10\%$ to $70\%$ of the decrease as seen in observations (see Fig. \ref{fig:wolf_B335}). This result appears to be in contrast to the RAT theory that causes a better alignment of the dust grains close to the radiation source. However, the degree of polarization depends not only on the alignment of dust grains but also, for instance, on the optical depth as seen in Fig. \ref{fig:mass_study}.
  
   \subsection*{Influence of the optical depth}
   \label{optical_depth_influence}
   
\begin{figure}
    \centering
    \includegraphics[width=\hsize]{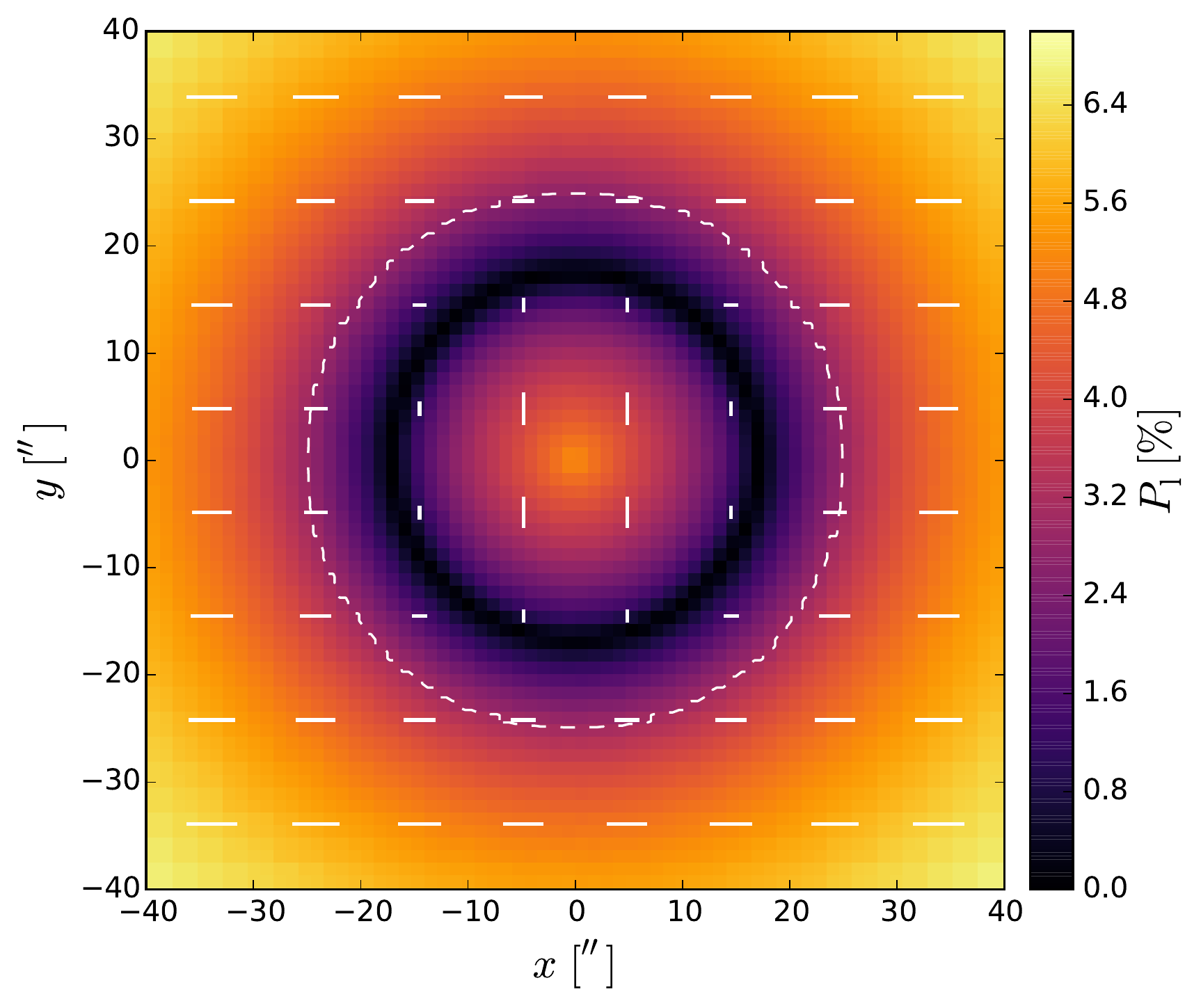}
    \caption{Polarization map of the thermal emission of our reference model with a total mass of $M_\mathrm{gas}=40~\mathrm{M_\odot}$ inside of $R=1.5\cdot10^4~\mathrm{AU}$ ($\lambda=450~\mathrm{\mu m}$). The image shows the innermost $80^{\prime\prime}\times 80^{\prime\prime}$ (corresponding to $8000~\mathrm{AU}\times8000~\mathrm{AU}$). The orientation of the white lines show the polarization angle and their length show the degree of polarization. The white dashed contour line shows the line-of-sight optical depth $\tau_\mathrm{LOS}=1$. The orientation of the magnetic field lines is parallel to the $y$-axis.}
    \label{fig:mass_40}
   \end{figure}

   Since the optical depth has a great influence on the polarized emission of Bok globules, we analyse how the optical depth influences linear polarization in general. For this, we have to adjust our reference model to feature optically thin, as well as optically thick line-of-sights. Therefore, we consider a total mass of $M_\mathrm{gas}=40~\mathrm{M_\odot}$ and simulate the thermal emission at $\lambda=450~\mathrm{\mu m}$. Figure \ref{fig:mass_40} illustrates a strong variation in the degree of polarization, which can be explained with the optical properties of elongated dust grains.
   
   \begin{figure*}
    \centering
     \includegraphics[width=0.9\hsize]{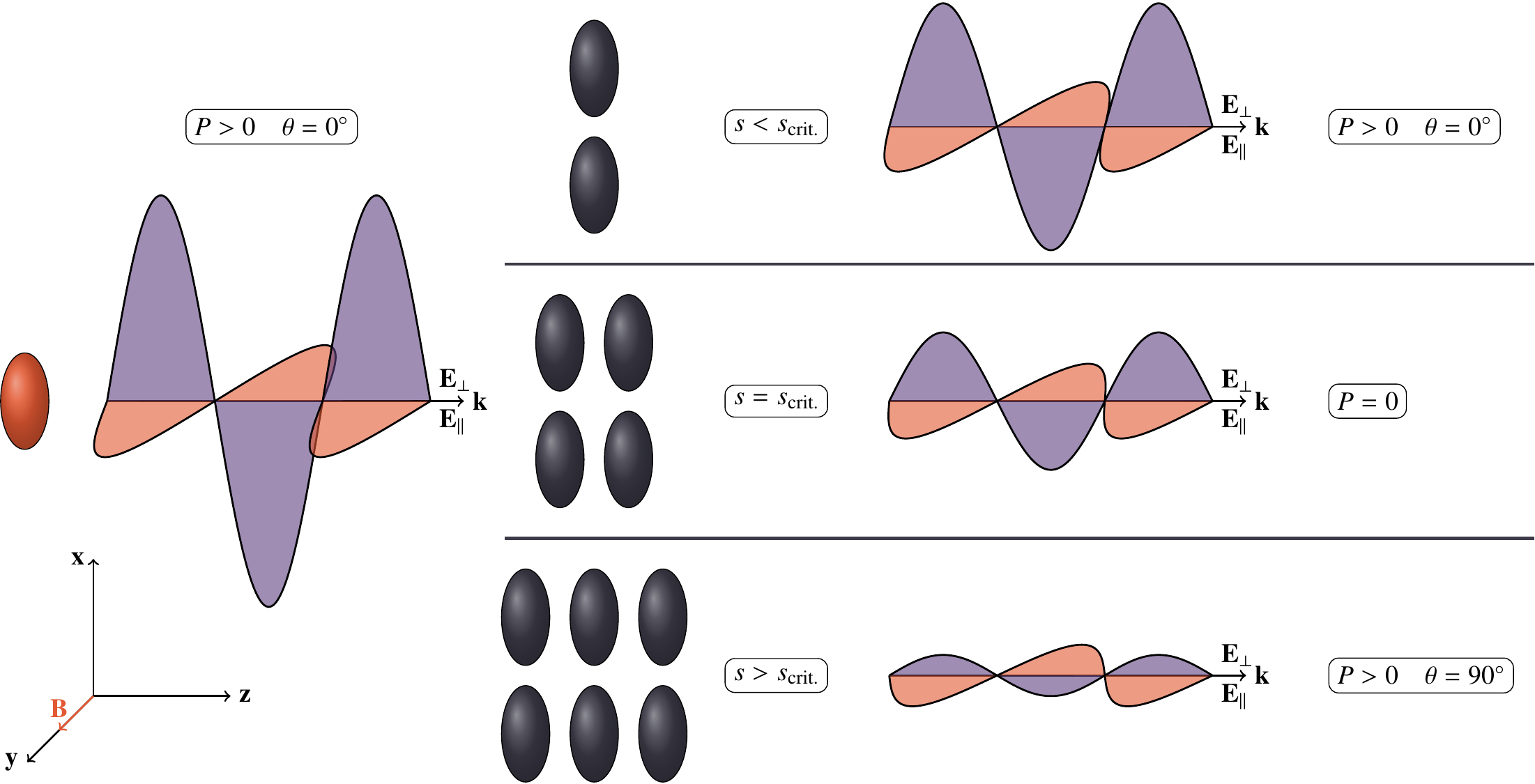}   
    \caption{Schematic illustration of the interaction between polarized thermal emission of an elongated prolate dust grain (orange) and their dichroic extinction (black). This illustration is also valid for oblate dust grains, because the key effect is the same. $\vec{E}$ is the electric field vector component parallel or perpendicular to the magnetic field $\vec{B}$. The critical distance $s_\mathrm{crit}$ is taken from Eq. \ref{eqn:s_crit}, $P_\mathrm{l}$ is the degree of polarization and $\theta$ is the polarization angle relative to the x-axis. The radiation of the absorbing dust grains (black) is not taken into account.}
    \label{fig:interaction_dichro}
   \end{figure*}

   The average thermal emission of an elongated dust grain can be split into a component of the longer $\perp$ and the shorter $\parallel$ axis:
   \begin{align}
    F_{\lambda}&=\frac{1}{3}\left(F_{\lambda,\perp}+2F_{\lambda,\parallel}\right),\\
    F_{\lambda,\perp}&=\frac{C^\mathrm{abs}_{\lambda,\perp}}{\sigma_\mathrm{geo}} B_\lambda\left(T_\mathrm{dust}\right),\\
    F_{\lambda,\parallel}&=\frac{C^\mathrm{abs}_{\lambda,\parallel}}{\sigma_\mathrm{geo}} B_\lambda\left(T_\mathrm{dust}\right).
   \end{align}
   Here, $B_\lambda\left(T_\mathrm{dust}\right)$ is the blackbody radiation, $\sigma_\mathrm{geo}$ is the projected geometrical surface area, and $C^\mathrm{abs}_{\lambda}$ is the absorption cross-section of the selected axis. After travelling a given distance, the radiation of elongated dust grains decreases as follows:
   \begin{align}
    F'_{\lambda,\perp}&=F_{\lambda,\perp}\exp\left(-\tau_\perp\right),\\
    F'_{\lambda,\parallel}&=F_{\lambda,\parallel}\exp\left(-\tau_\parallel\right),
   \end{align}
   where the optical depth $\tau$ is defined as
   \begin{align}
    \tau_\perp&=\int n_\mathrm{dust}C^\mathrm{ext}_{\lambda,\perp} \mathrm{ds},\label{eqn:tau_1}\\
    \tau_\parallel&=\int n_\mathrm{dust}C^\mathrm{ext}_{\lambda,\parallel} \mathrm{ds}.\label{eqn:tau_2}
   \end{align}
    Here, $C^\mathrm{ext}_{\lambda}$ is the extinction cross-section parallel or perpendicular to the shorter axis of the first dust grain and $n_\mathrm{dust}$ is the dust grain number density. If we assume that each dust grain has the same orientation, composition, and extent in the model space (constant cross-sections), $F'_{\lambda}$ can be derived as follows:
   \begin{align}
    F'_{\lambda,\perp}&=\frac{C^\mathrm{abs}_{\lambda,\perp}}{\sigma_\mathrm{geo}} B_\lambda\left(T_\mathrm{dust}\right)\exp\left(-n_\mathrm{dust}C^\mathrm{ext}_{\lambda,\perp}s\right),\label{eqn:first}\\
    F'_{\lambda,\parallel}&=\frac{C^\mathrm{abs}_{\lambda,\parallel}}{\sigma_\mathrm{geo}} B_\lambda\left(T_\mathrm{dust}\right)\exp\left(-n_\mathrm{dust}C^\mathrm{ext}_{\lambda,\parallel}s\right).\label{eqn:last}
   \end{align}
   With Eqs. \ref{eqn:first} and \ref{eqn:last}, we can find the critical distance at which the polarization is zero ($F'_{\lambda,\perp}=F'_{\lambda,\parallel}$).
   \begin{equation}
    s_\mathrm{crit}=\frac{1}{n_\mathrm{dust}\left(C^\mathrm{ext}_{\lambda,\perp}-C^\mathrm{ext}_{\lambda,\parallel}\right)} \ln\left(\frac{C^\mathrm{abs}_{\lambda,\perp}}{C^\mathrm{abs}_{\lambda,\parallel}}\right)
    \label{eqn:s_crit}
   \end{equation}
   The work of \cite{reissl_tracing_2014} shows a similar analysis, but focuses on the sign change of the Stokes vector Q-component instead of the critical distance.
   
   If we use the cross-sections of our dust grain model, $s_\mathrm{crit}$ corresponds to an optical depth in the order of about one but always lower than one. The quantity $s_\mathrm{crit}n_\mathrm{dust}$ becomes unity in the limit of spherical dust grains, which is expected from analytical considerations (Poisson distribution).

   Therefore, we can distinguish three different cases for the polarized emission of an elongated dust grain (see Fig. \ref{fig:interaction_dichro}). If  $s<s_\mathrm{crit}$, the dichroic emission dominates the degree of polarization. With $s=s_\mathrm{crit}$, the dichroic emission and extinction compensate each other and the degree of polarization is zero. In the case of $s>s_\mathrm{crit}$, the dichroic extinction increases the degree of polarization again and rotates the polarization angle by $90^\circ$. The precise value of $\tau_\mathrm{crit}$ depends on the distribution, orientation, and geometrical shape of the dust grains.

   A significant decrease in the degree of polarization is therefore achievable with a line-of-sight optical depth in the order of one.

  \subsection*{Dust grain size}
  \label{dust_grain_size}

  \begin{figure}[ht!]
   \centering
    \includegraphics[width=\hsize]{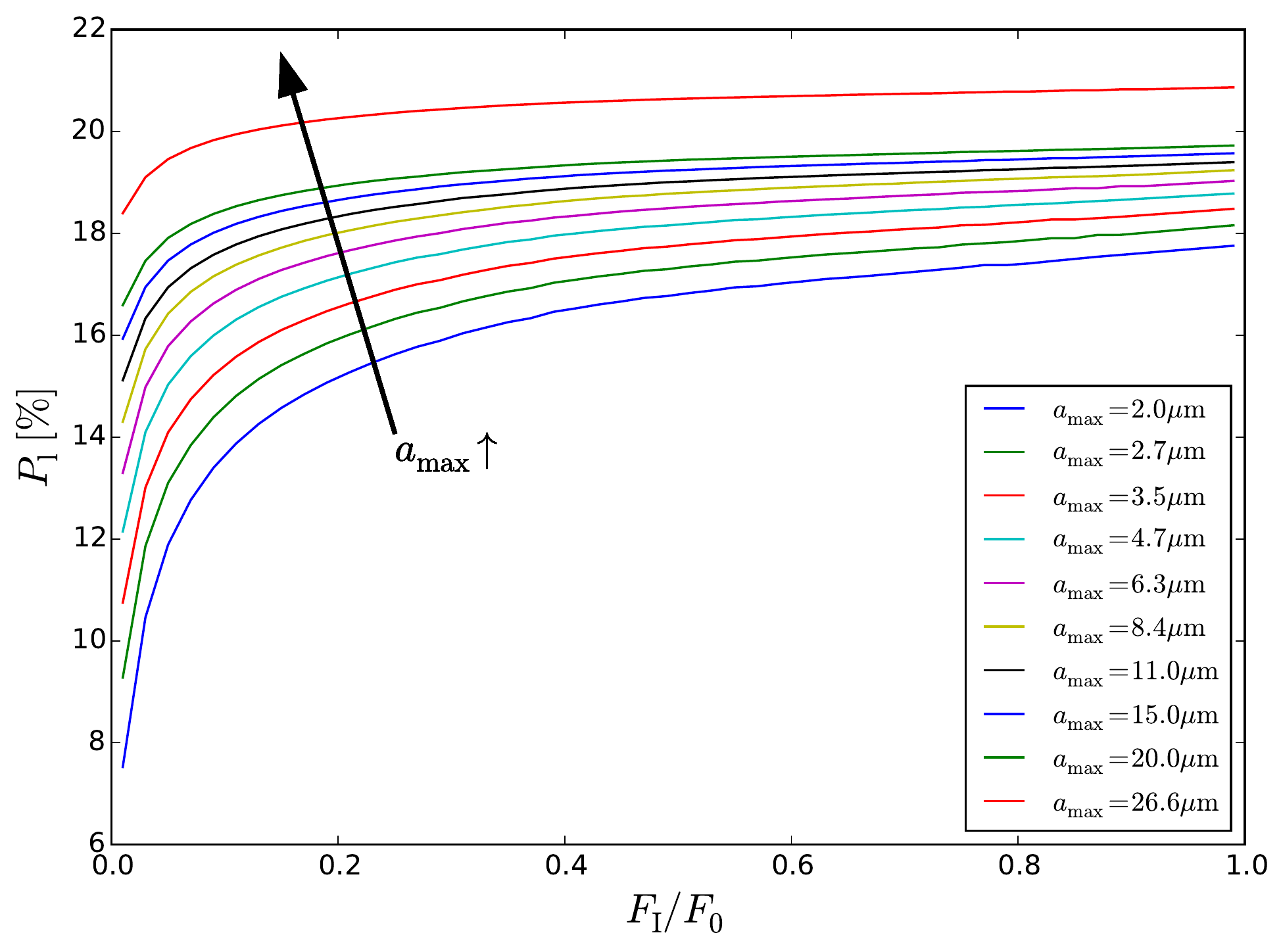}
    \includegraphics[width=\hsize]{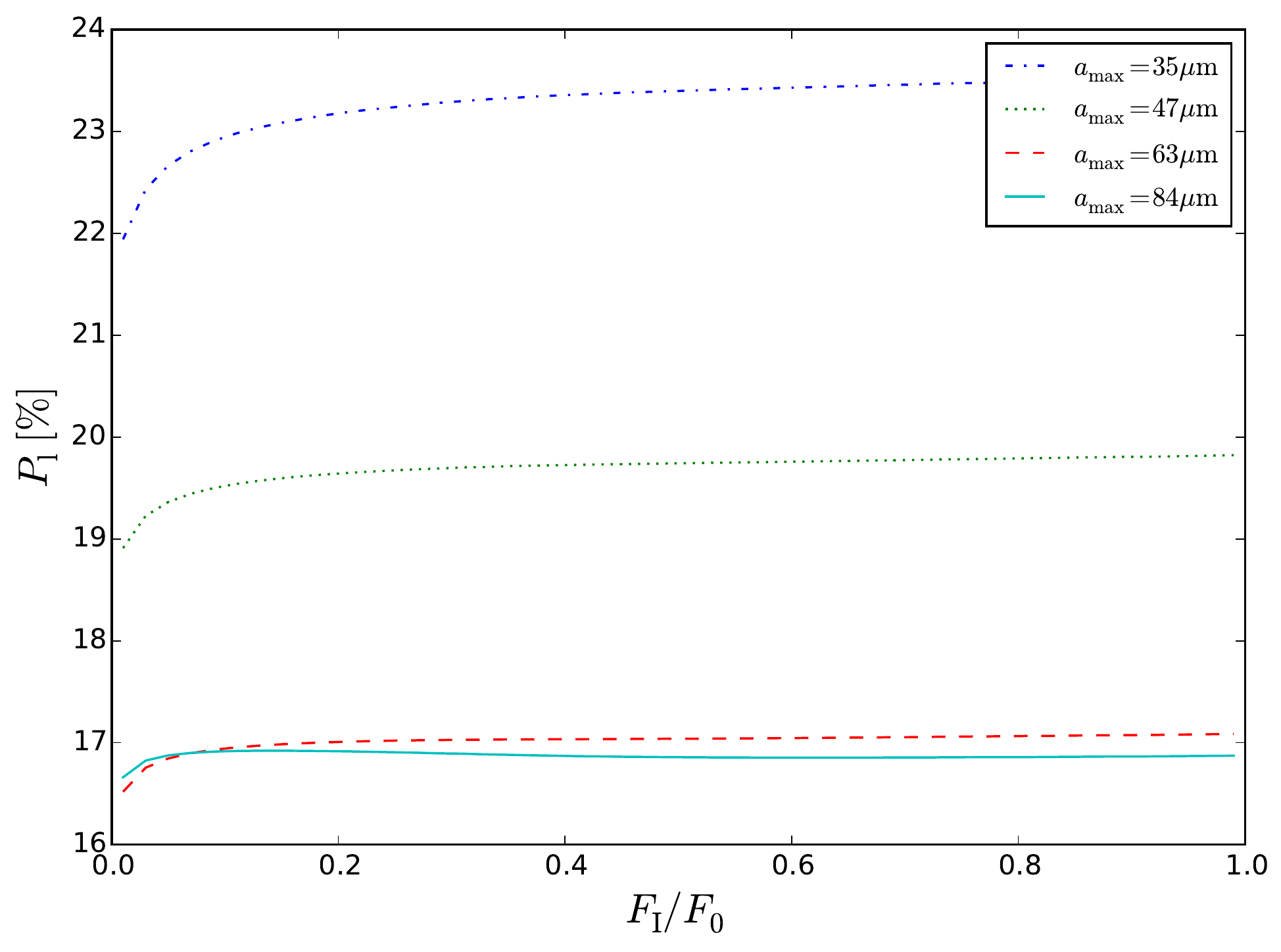}
    \includegraphics[width=\hsize]{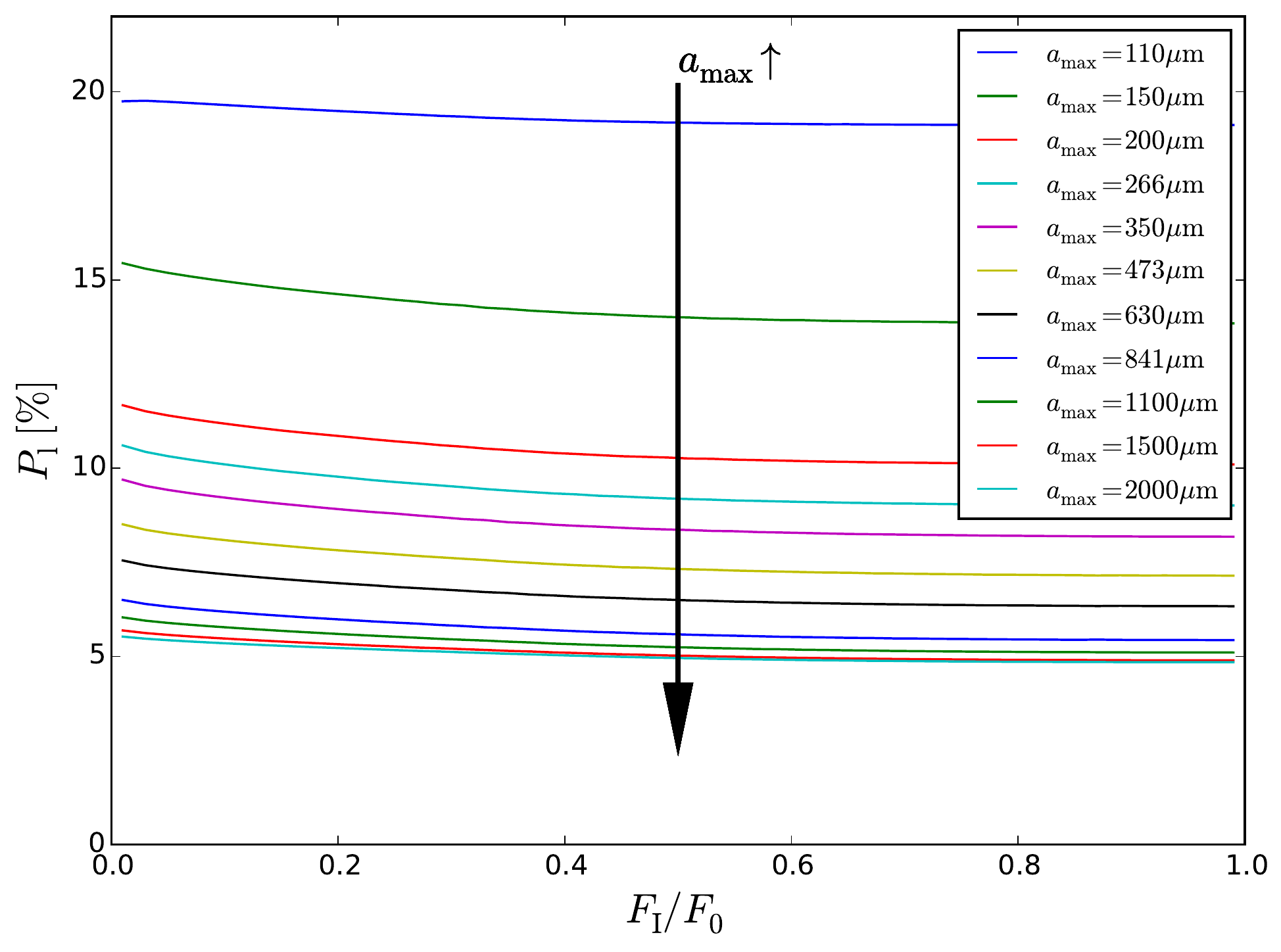}

   \caption{Distribution of the mean degree of polarization $P_\mathrm{l}$ vs. the intensity $F_\mathrm{I}$ across the Bok globule models. The intensity $F_\mathrm{I}$ is normalized to the maximum intensity $F_0$ of the respective Bok globule. The Bok globule models have different maximum dust grain sizes from $a_\mathrm{max}=2~\mathrm{\mu m}$ to $a_\mathrm{max}=2000~\mathrm{\mu m}$. The distributions are split in three images to outline the different behaviour of the polarization degree.}
   \label{fig:dust_grain_size}
  \end{figure}

  Besides the dust mass, the dust grain size is also expected to have a great impact on several characteristics of a Bok globule. For instance, the absorption and extinction cross-sections are highly dependent on the dust grain size (distribution) and influence the optical depth (see Eqs. \ref{eqn:tau_1} and \ref{eqn:tau_2}). The radiative torque alignment is also sensitive to the size of the dust grains. We therefore investigate the influence of different dust grain size distributions on the behaviour of the degree of polarization in Bok globules. For this purpose, we simulate our reference model using different maximum dust grain sizes $a_\mathrm{max}\in\{2,\dots,200,\dots,2000\}~\mathrm{\mu m}$. These maximum dust grain sizes correspond to the following optical depths:
  \begin{align*}
   a_\mathrm{max}&=2~\mathrm{\mu m}: \quad \tau_{850~\mathrm{\mu m}}=3\cdot10^{-3}\quad (A_V=90)\\
   a_\mathrm{max}&=200~\mathrm{\mu m}: \quad \tau_{850~\mathrm{\mu m}}=0.12\quad (A_V=10)\\
   a_\mathrm{max}&=2000~\mathrm{\mu m}: \quad \tau_{850~\mathrm{\mu m}}=0.07\quad (A_V=3).
  \end{align*}
  Here, the optical depth is derived for the distance between the centre and the boundary of our Bok globule model ($M_\mathrm{gas}=8~\mathrm{M_\odot}$, $\lambda=850~\mathrm{\mu m}$).

  As illustrated in Fig. \ref{fig:dust_grain_size} (top), the distribution of the degree of polarization vs. intensity shows between $a_\mathrm{max}\sim2~\mathrm{\mu m}$ and $a_\mathrm{max}\sim30~\mathrm{\mu m}$ a maximum in the high intensity central region and a significant decrease towards the outer low intensity parts of the Bok globule. Even up to $a_\mathrm{max}\sim80~\mathrm{\mu m}$, no polarization hole can be seen in the profile (see Fig. \ref{fig:dust_grain_size}, middle). Polarization holes only occur with dust grains above $a_\mathrm{max}\sim80~\mathrm{\mu m}$ (see Fig. \ref{fig:dust_grain_size}, bottom). Furthermore, these threshold values of the maximum dust grain size hardly depend on the total Bok globule mass.

  As illustrated in Fig. \ref{fig:dust_grain_size} (bottom), dust grain size distributions with $a_\mathrm{max}>80~\mathrm{\mu m}$ generally decrease the degree of polarization of the Bok globule model with increasing size. According to the work of \cite{weidenschilling_coagulation_1994}, the densest regions in Bok globules are expected to consist of larger dust grains than the less dense regions. For instance, with dust grains up to a few $100~\mathrm{\mu m}$ in the centre of the Bok globule model and grains smaller than $10~\mathrm{\mu m}$ in the surroundings, a difference of the degree of polarization of $P\sim10\%$ can be achieved.

  \subsection*{Dust grain composition}
  \label{dust_grain_composition}

  \begin{figure}
   \centering
   \includegraphics[width=\hsize]{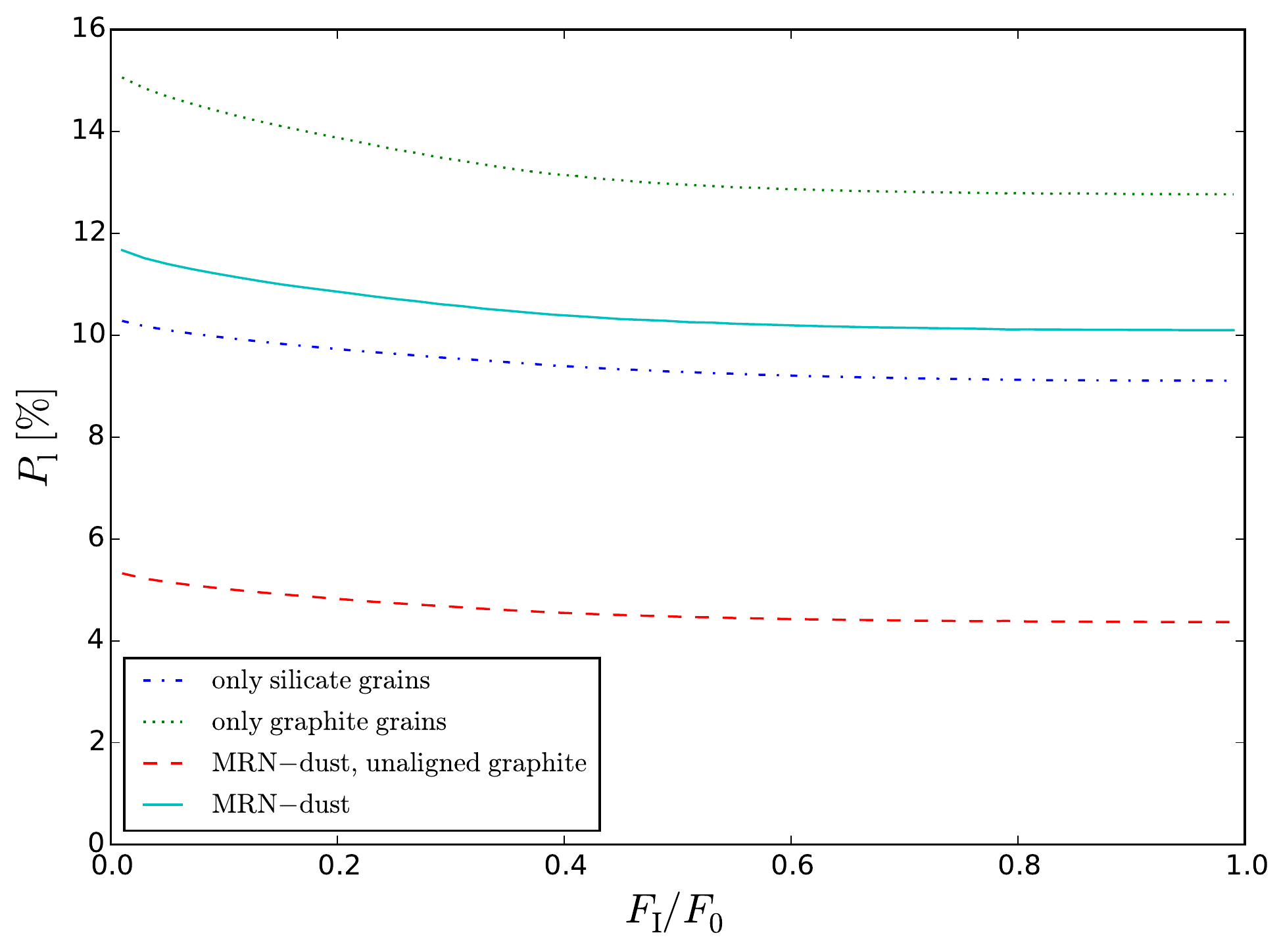}
   \caption{Distribution of the mean degree of polarization $P_\mathrm{l}$ vs. the intensity $F_\mathrm{I}$ across the Bok globule models. The intensity $F_\mathrm{I}$ is normalized to the maximum intensity $F_0$ of the respective Bok globule. The dust grains in the Bok globule models consist only of silicate, graphite, MRN-dust or MRN-dust with unaligned graphite grains.}
   \label{fig:components}
  \end{figure}

  The composition of dust grains in Bok globules is not exactly known and may be different in the Bok globule core and envelope \citep{olofsson_mid-infrared_2011}. As used in the reference model, the MRN-dust composition is a popular choice which is in agreement with many observations. However, the ratio between silicate and graphite grains does not need to be the same in every part of a Bok globule. To investigate the influence of the relative abundance on the degree of polarization, we simulate the reference model with MRN-dust and dust grains that consist of only one component. As illustrated in Fig. \ref{fig:components}, graphite grains cause an overall higher degree of polarization and a slightly stronger decrease in the degree of polarization than silicate grains.
  
  \cite{olofsson_mid-infrared_2011} mention that the dust phase in the centre of the Bok globule B335 consists of much more graphite grains than the surroundings. This may also occur in other Bok globules and increase in the degree of polarization towards the higher amount of graphite grains. In that case, mechanisms which cause polarization holes in Bok globules need to compensate for the influence of the higher amount of aligned graphite grains. However, this impact of the graphite grains is only true as long as they align with the magnetic field, which is proposed by the works of \cite{clayton_dust_2003} and \cite{draine_infrared_2007}. In contrast, ({\color{blue}\citealt{mathis_alignment_1986}; \citealt{lazarian_tracing_2007}; \citealt{andersson_interstellar_2015}}) mentioned that graphite grains do not align with the magnetic field. If the latter statement is correct, the degree of polarization decreases in regions with a higher amount of graphite grains, which supports polarization holes. For instance, taking MRN-composition in the central region of the Bok globule model and pure silicate in the surroundings into account, the decrease in the degree of polarization amounts to $\sim5\%$ (see Fig. \ref{fig:components}).

  \subsection*{Radiative torque alignment}
  \label{RAT}

  \begin{figure}
   \centering
   \includegraphics[width=\hsize]{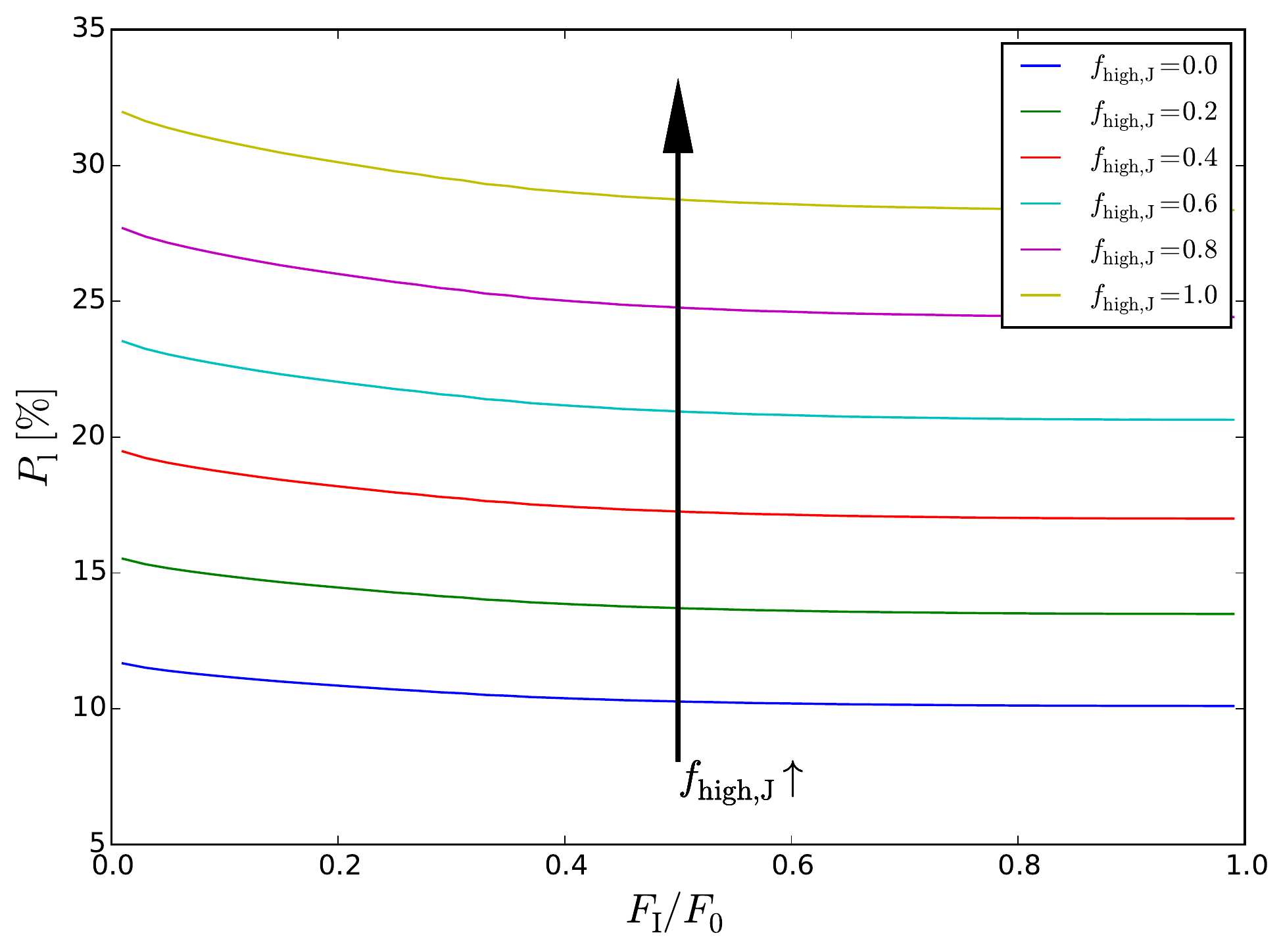}
   \caption{Distribution of the mean degree of polarization $P_\mathrm{l}$ vs. the intensity $F_\mathrm{I}$ across the Bok globule models. The intensity $F_\mathrm{I}$ is normalized to the maximum intensity $F_0$ of the respective Bok globule. The Bok globule models have different values of $f_\mathrm{high,J}$ from $0.0$ to $1.0$ in $0.2$ steps.}
   \label{fig:polarization_fhighj}
  \end{figure}

  As described in Sect. \ref{radiative_transfer}, the efficiency of the radiative torque alignment depends strongly on the free parameter $f_\mathrm{high,J}$. To estimate a potential influence on the optical depth effect and the degree of polarization in general, we simulate our reference model using values of $f_\mathrm{high,J}$ from $0$ to $1.0$ in $0.2$ steps. As predicted, an increase in $f_\mathrm{high,J}$ causes a great increase in the degree of polarization (see Fig. \ref{fig:polarization_fhighj}). Nevertheless, the relative decrease in the degree of polarization does not change through variations in $f_\mathrm{high,J}$. 

  Polarimetric observations of Bok globules show a degree of polarization of $P{\sim}15\%$ in their outer regions \citep[see e.g.][]{henning_measurements_2001}. We therefore estimate that the mean value of $f_\mathrm{high,J}$ is less than $0.4$ in Bok globules, which is in agreement with the results of \cite{hoang_radiative_2008}.

  As mentioned in Sect. \ref{radiative_transfer}, a constant value of $f_\mathrm{high,J}$ in the whole Bok globule is not expected in reality. A spatial variation in $f_\mathrm{high,J}$ is more suitable for taking variations in the temperature and density distribution into account. Because of the high temperature and density in the centre of Bok globules, a lower ratio of fast to low spinning dust grains is reasonable. As illustrated in Fig. \ref{fig:polarization_fhighj}, a lower value of $f_\mathrm{high,J}$ in the dense regions of Bok globules would decrease the degree of polarization. This effect is similar to the explanation of increased disalignment owing to higher temperature and density, but can now be explained by fewer dust grains, which rotate at high angular momentum considering the RAT theory.

  \subsection*{Bok globules with no central protostar}
  \label{no_central_protostar}
  
  Some Bok globules have starless cores instead of one or more central protostars \citep{launhardt_looking_2010}. With our present results, we discuss the effect of a missing central protostar on the degree of polarization.
  
  The critical distance $s_\mathrm{crit}$ of the optical depth effect depends mainly on the alignment efficiency in the most massive regions (see Influence of the optical depth in Sect. \ref{optical_depth_influence}). Without a central protostar, these regions, which are mainly the core of a Bok globule, are not well irradiated by an external radiation source and, therefore, not efficiently aligned by the RAT mechanism. As a result, the emission of the core is not strongly polarized and the aforementioned optical depth effect is rather weak. However, since the dust grains in the core are not well aligned, the degree of polarization is still decreasing towards the dense regions of Bok globules. Thus, we expect that the polarized emission of Bok globules with or without a central protostar show the same trend, but based on different underlying physical effects.
  
  \subsection*{Further investigations}
  \label{further_investigations}

\begin{figure}[htpb]
  \centering
   \includegraphics[width=\hsize]{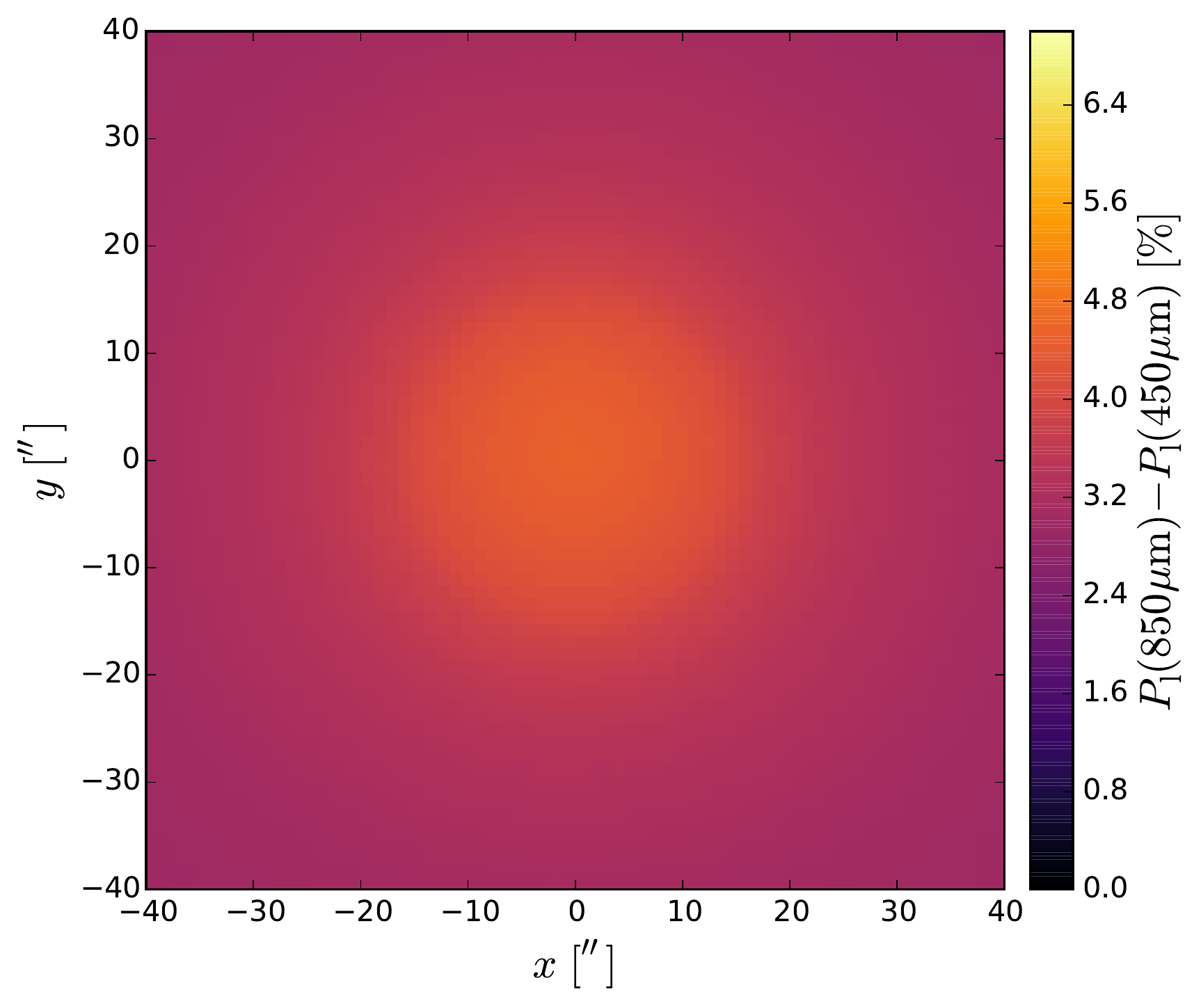}
   \includegraphics[width=\hsize]{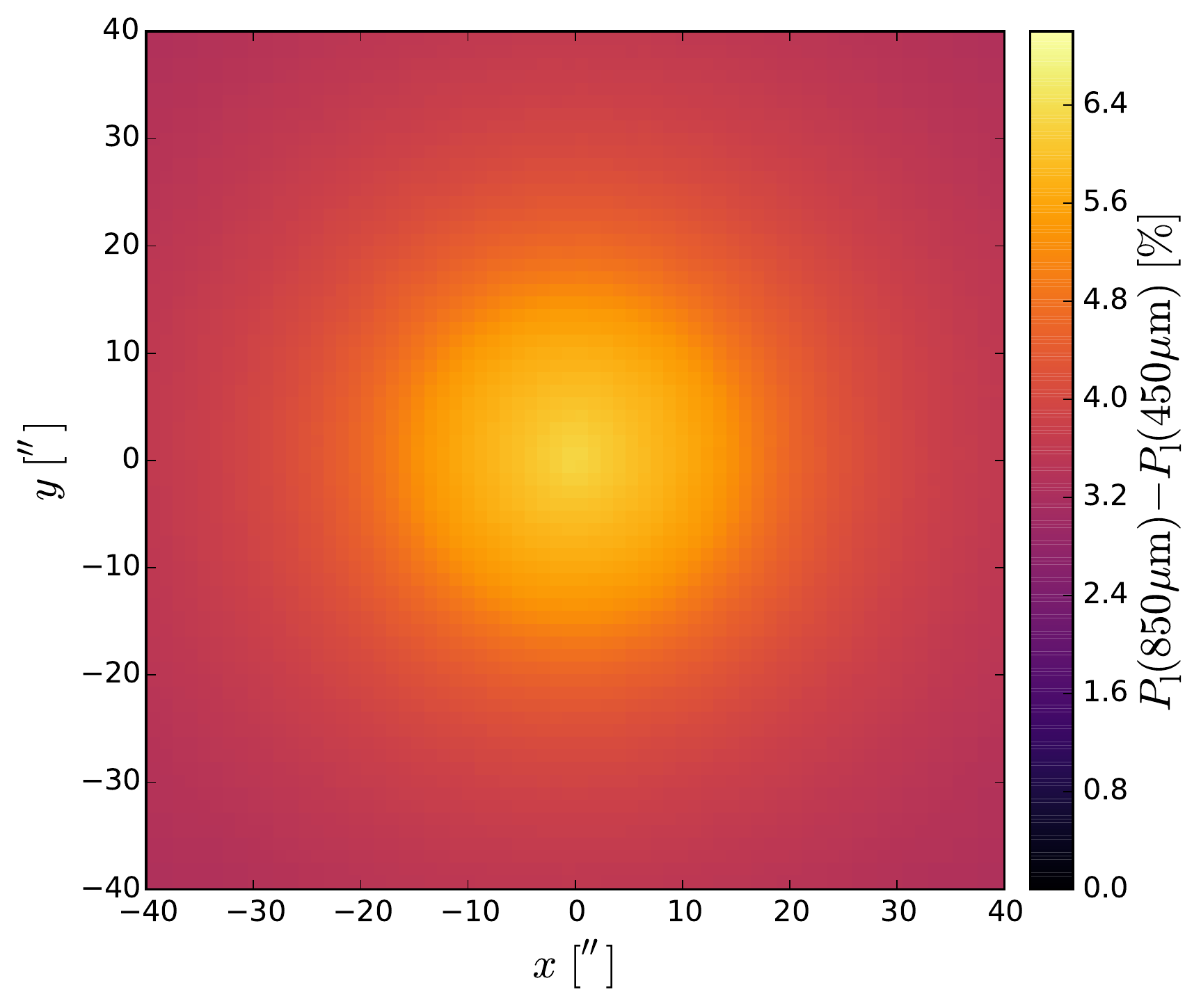}
   \includegraphics[width=\hsize]{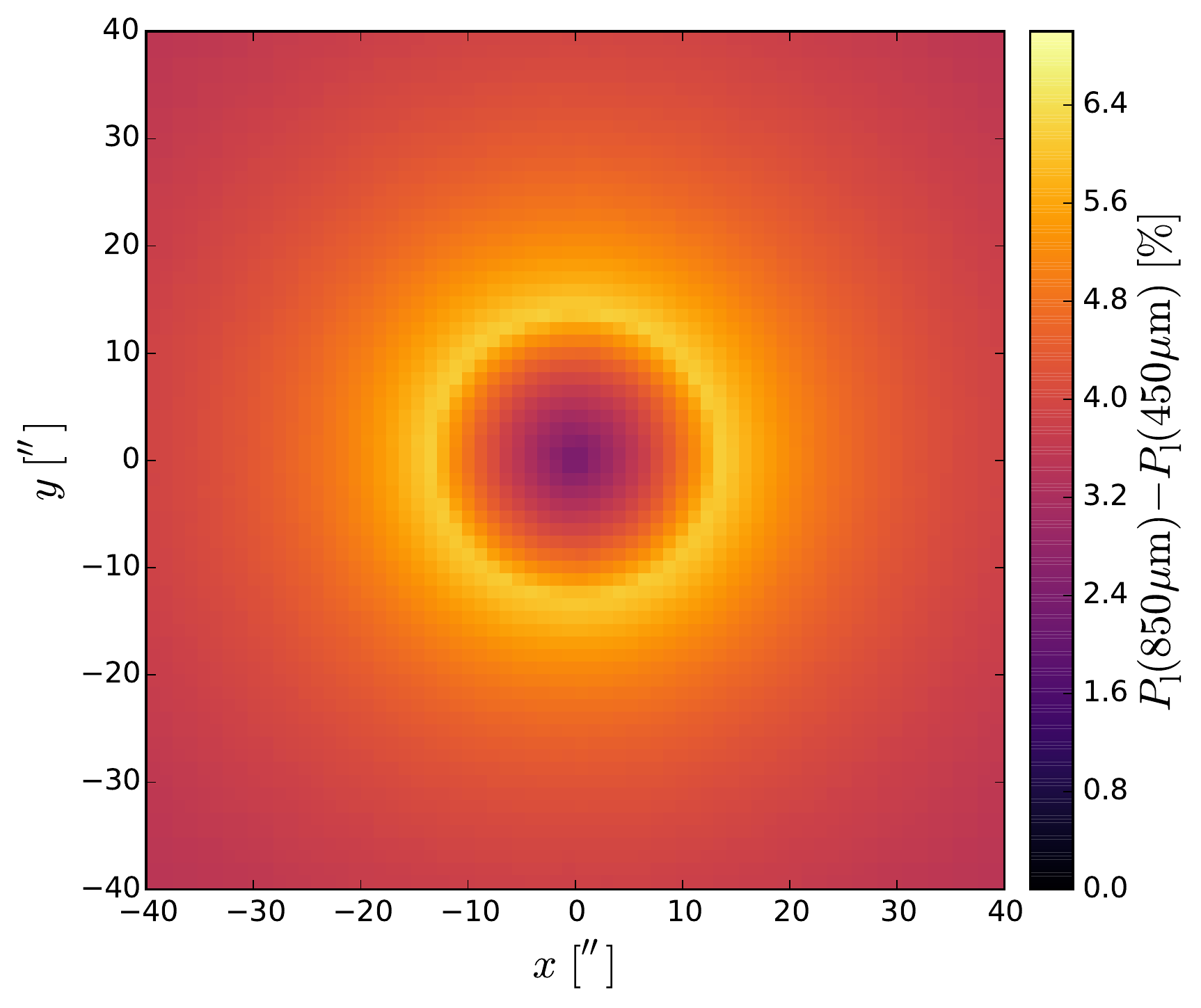} 

  \caption{Difference images between the thermal emission linear polarization maps of our reference model observed at $\lambda=450~\mathrm{\mu m}$ and $\lambda=850~\mathrm{\mu m}$ (top: $M_\mathrm{gas}=12~\mathrm{M_\odot}$, middle: $M_\mathrm{gas}=24~\mathrm{M_\odot}$, bottom: $M_\mathrm{gas}=32~\mathrm{M_\odot}$). The images show the innermost $80^{\prime\prime}\times 80^{\prime\prime}$ (corresponding to $8000~\mathrm{AU}\times8000~\mathrm{AU}$).}
 \label{fig:observation_test}
\end{figure}

  At shorter wavelengths (e.g. $\lambda=450~\mathrm{\mu m}$), the line-of-sight optical depth increases which influences the degree of polarization as described in Sect. \ref{optical_depth_influence} (see Fig. \ref{fig:mass_40}). Observations of Bok globules at shorter wavelengths could be used to constrain the impact of the optical depth effect on the degree of polarization. The images in Fig. \ref{fig:observation_test} show the difference between the polarization maps at $\lambda=450~\mathrm{\mu m}$ and $\lambda=850~\mathrm{\mu m}$ of our reference model using different total masses. A large difference of the linear polarization can be related to a great impact of the optical depth effect at $\lambda=850~\mathrm{\mu m}$ (Fig. \ref{fig:observation_test}, middle). This impact is even greater if the degree of polarization is in the same order of magnitude at both wavelengths, but has its orientation shifted by $90^\circ$ (Fig. \ref{fig:observation_test}, bottom).  
  
  A possible observation could be performed with SCUBA-2/JCMT at $\lambda=450~\mathrm{\mu m}$. From simulations of our reference model, we obtain a required sensitivity of $\Delta F'_\mathrm{I}\sim19~\mathrm{mJy/beam}$ to resolve the important inner region of the Bok globule \citep{wardle_linear_1974, reduce_pol}:
  \begin{equation}
   \Delta F'_\mathrm{I}\approx \frac{1}{2}PF_\mathrm{I}\cdot\frac{\Delta\gamma}{29^\circ}, \qquad \Delta\gamma\sim10^\circ.
  \end{equation}
  Here, $\Delta\gamma$ is the required resolution of the polarization angle, $F_\mathrm{I}$ the Intensity and $P_\mathrm{l}$ the degree of polarization. Taking the Daisy map (field of view of $\sim3^{\prime}$) and standard values for weather and airmass of the SCUBA-2 Integration Time Calculator \citep{holland_scuba-2:_2013} into consider, we derive an observation time of about two hours and 15 minutes.

 \section{Conclusions}
 \label{conclusions}

  We investigated the influence of various physical conditions and quantities on the occurrence of polarization holes in Bok globules and the degree of polarization in general. Under selected circumstances, we found that variations in the optical depth, the dust grain size, and the dust grain composition significantly influence the degree of polarization in Bok globules.
  
  The conditions under which the optical depth significantly decreases the degree of polarization are summarised as follows:
  \begin{enumerate}
   \item \textbf{Dust grain growth to submm/mm size\\} 
   The absorption cross-section of dust grains grown to $80~\mathrm{\mu m}$ or less (spectral index of $>3$ or $\beta>1$) is too low, in combination with typical column densities of Bok globules, to cause a significant decrease in the degree of polarization. However, dust grains grown to submm/mm size are able to decrease the polarization and are consistent with the spectral index derived from observed fluxes of various Bok globules at $\lambda=850~\mathrm{\mu m}$ and $\lambda=1.3~\mathrm{mm}$ ($\tau<3$ or $\beta<1$).
   \item \textbf{Hydrogen column density of $\boldsymbol{N_\mathrm{H}\sim10^{28}m^{-2}}$}\\
   Given the dust grain growth above, the column density needs to be in the order of magnitude of about $10^{28}m^{-2}$ to decrease the degree of polarization by up to $\Delta P\sim10~\%$. This corresponds to a total mass of some $10~\mathrm{M_\odot}$ inside of $1.5\cdot10^{4}~\mathrm{AU}\times1.5\cdot10^{4}~\mathrm{AU}$ which is in agreement with masses found for Bok globules.
  \end{enumerate}
  The following physical quantities and conditions are also able to significantly decrease the degree of polarization in Bok globules:
  \begin{enumerate}
   \item \textbf{Larger dust grains limited to dense regions\\} 
   A gradient of the dust grain size towards submm/mm grains in the dense regions of Bok globules is able to decrease the degree of polarization by up to $\Delta P\sim10~\%$.
   \item \textbf{Unaligned graphite grains accumulated in dense regions\\} 
   The combination of graphite grains that are not aligned with the magnetic field and a higher amount of graphite grains in dense regions of Bok globules is able to decrease the degree of polarization by up to $\Delta P\sim5\%$.
   \item \textbf{Less efficient grain alignment in the central region\\} 
   If the higher density and temperature in the dense regions of Bok globules reduce the amount of dust grains that rotate at high angular momentum, the degree of polarization decreases by up to $\Delta P\sim10~\%$ by applying the RAT theory and assuming $f_\mathrm{high,J}<0.4$.
   \item \textbf{No central stellar heating source\\} 
   In Bok globules without a central stellar heating source, the degree of polarization is expected to decrease similarly to Bok globules with a heating source. This is caused by the low alignment efficiency in the central dense core, which is due to the weak irradiation from external radiation sources.
  \end{enumerate}
  In addition to increased disalignment and twisted magnetic field lines, we found that the above conditions (optical depth, dust grain size, and dust grain composition) provide an explanation for the significant decrease in linear polarization towards dense regions of Bok globules. Given typical physical conditions in Bok globules, several effects are likely to cause the observed polarization holes together.

 \begin{acknowledgements}
       Part of this work was supported by the German
       Deut\-sche For\-schungs\-ge\-mein\-schaft, DFG\/ project number WO 857/12-1.
 \end{acknowledgements}

\nocite{reissl}
\bibliographystyle{aa}
\bibliography{bibtex}

\end{document}